\documentclass[showpacs,aps,graphicx,twocolumn]{revtex4}%
\usepackage{graphicx}

\begin{document}
\title{Fake-signal-and-cheating attack on quantum secret sharing}
\author{ Fu-Guo Deng,$^{1,2,3}$ Xi-Han Li,$^{1,2}$ Pan Chen,$^4$ Chun-Yan Li,$^{1,2}$
and Hong-Yu Zhou$^{1,2,3}$}
\address{$^1$ The Key Laboratory of Beam Technology and Material
Modification of Ministry of Education, Beijing Normal University,
Beijing 100875,
People's Republic of China\\
$^2$ Institute of Low Energy Nuclear Physics, and Department of
Material Science and Engineering, Beijing Normal University,
Beijing 100875, People's Republic of China\\
$^3$ Beijing Radiation Center, Beijing 100875,  People's Republic of
China\\
$^4$  Key Laboratory For Quantum Information and Measurements and
Department of Physics, Tsinghua University, Beijing 100084, People's
Republic of China}
\date{\today }

\begin{abstract}
We present a fake-signal-and-cheating attack strategy for the
dishonest agent in quantum secret sharing (QSS) to steal the
information of the other agents' fully and freely. It is found that
almost all the QSS protocols existing, such as the two famous QSS
protocols, the Hillery-Bu$\check{z}$ek-Berthiaume [Phys. Rev. A
\textbf{59}, 1829 (1999)] and the Karlsson-Koashi-Imoto  [Phys. Rev.
A \textbf{59}, 162 (1999)], can be eavesdropped freely if the
process for the eavesdropping check is accomplished with the
cooperation of the dishonest agent. He can sends a fake signal to
the other agents after intercepting the original photons and storing
them. His action can be hidden with entanglement swapping and
cheating when the photons are chosen as the samples for checking
eavesdropping. Finally, we present a possible improvement of these
QSS protocols' security with decoy photons.

\end{abstract}
 \maketitle

Quantum information, an ingenious application of quantum mechanics
within the field of information has attracted a lot of attention
\cite{book,Gisinqkd}. In particular almost all the branches of
quantum communication have been developed quickly since the original
protocol was proposed by Bennett and Brassard (BB84) \cite{BB84} in
1984, such as quantum key distribution (QKD)
\cite{Ekert91,BBM92,Hwang,ABC}, quantum secure direction
communication \cite{QSDC0,QSDC1,two-step}, quantum secret sharing
\cite{HBB99,KKI,Bandyopadhyay,Karimipour,Nascimento,Gottesman,cpyang,dengQSSentanglement,longqss,guoqss,dengsinglephoton,zhangPLA,zhanglm,yanpra,
cleve,Peng,dengmQSTS,dengpra,TZG,AMLance}, and so on. QKD provides a
secure way \cite{Gisinqkd} for creating a private key between two
remote parties, say Bob and Charlie. With a private key, most of the
goals in classical secure communication can be accomplished. For
example, a classical secret message can be transmitted securely by
using classical one-time pad crypto-system with a private key.

In a secret sharing \cite{Blakley}, a boss, say Alice, has two
agents, Bob and Charlie who are at remote places. Alice hopes that
the agents can carry out her instruction (the message $M_A$), but
she doubts that one of them (not more than one) may be dishonest and
he will do harm to her belongings if the dishonest one can deal with
it independently. Moreover, Alice does not know who the dishonest
one is. For the security of the secret message $M_A$ , Alice splits
it into two pieces, $M_B$ and $M_C$, and then she sends them to Bob
and Charlie, respectively. The two agents can read out the message
$M_A=M_B\oplus M_C$ if and only if they cooperate, otherwise none of
them can obtain an useful information. QSS is the generalization of
classical secret sharing into quantum scenario and has progressed
quickly in recent years
\cite{HBB99,KKI,Bandyopadhyay,Karimipour,Nascimento,Gottesman,cpyang,dengQSSentanglement,longqss,guoqss,dengsinglephoton,zhangPLA,zhanglm,yanpra,
cleve,Peng,dengmQSTS,dengpra,TZG,AMLance}. It provides a secure way
for sharing not only  classical information
\cite{HBB99,KKI,Bandyopadhyay,Karimipour,Nascimento,Gottesman,cpyang,dengQSSentanglement,longqss,guoqss,dengsinglephoton,zhangPLA,zhanglm,yanpra},
but also quantum information \cite{cleve,Peng,dengmQSTS,dengpra}. In
the latter, the sender, Alice will send an unknown quantum state to
her $m$ agents and one of them can recover it with the help of the
others \cite{dengmQSTS}. Now, QSS has also been studied in
experiment \cite{TZG,AMLance}. In general, QSS is by far more
complex than QKD as the dishonest agent, a powerful eavesdropper,
has the chance to hide his eavesdropping with cheating.

Almost all the QSS protocols existing can be attributed to one of
the two types, the collective eavesdropping-check one and the
individual eavesdropping-check one. The feature of the QSS protocols
based on a collective eavesdropping check is that the process of the
eavesdropping check can be completed only with the cooperation of
the dishonest agent. That is, the boss Alice should require the
dishonest agent to publish his outcomes of the samples for checking
eavesdropping or his operations on them. Otherwise, she and the
honest one cannot accomplish the process. The typical models are the
two famous QSS protocols, the two pioneers, the
Hillery-Bu$\check{z}$ek-Berthiaume 1999 (HBB99) \cite{HBB99} and the
Karlsson-Koashi-Imoto (KKI) \cite{KKI}. In contrast, in the
individual eavesdropping-check QSS protocols, the eavesdropping
check between the boss Alice and the honest agent need not resort to
the information published by the dishonest one. In other words,
Alice and the honest agent can analyze the error rate of their
samples independent on the dishonest one. Typical such QSS protocols
are the ones presented in Refs. \cite{guoqss,dengsinglephoton}. As
the security of a QSS protocol is based on the error rate analysis
of the samples chosen by the parties of the communication, the
dishonest agent can obtain all the information about other agents'
if his eavesdropping does not introduce the errors in the samples.
In fact, almost all the QSS protocols
\cite{HBB99,KKI,Bandyopadhyay,Karimipour,Nascimento,Gottesman,cpyang,dengQSSentanglement,longqss,zhangPLA,zhanglm,yanpra}
based on the collective eavesdropping check can be eavesdropped
fully and freely with a cheat if he has the capability of performing
entanglement swapping \cite{entanglementswapping} and storing the
quantum states.

In this paper, we will introduce a fake-signal-and-cheating attack
on the QSS protocols
\cite{HBB99,KKI,Bandyopadhyay,Karimipour,Nascimento,Gottesman,cpyang,dengQSSentanglement,longqss,zhangPLA,zhanglm,yanpra}
based on the collective eavesdropping check. It will be shown that
the dishonest agent can steal all the information of the other
agents' freely and fully with entanglement swapping and cheating.
Moreover, we present a possible improvement of those QSS protocols'
security with some decoy photons. We limit our discuss in the two
famous QSS protocols, HBB99 \cite{HBB99} and KKI \cite{KKI}. For the
other protocols, the principle is same as them  with or without a
little modification.

Same as that in HBB99 QSS protocol \cite{HBB99}, the
Greenberger-Horne-Zeilinger (GHZ) triplet that Alice prepares is in
the state
\begin{eqnarray}
\vert \Psi\rangle&=&\frac{1}{\sqrt{2}}(\vert 000\rangle + \vert 111\rangle)_{ABC}, \nonumber\\
&=& \frac{1}{\sqrt{2}}[(\vert +x\rangle_A\vert +x\rangle_B + \vert
-x\rangle_A\vert -x\rangle_B)\vert +x\rangle_C \nonumber\\
&& + (\vert +x\rangle_A\vert -x\rangle_B + \vert -x\rangle_A\vert
+x\rangle_B)\vert -x\rangle_C \nonumber\\
&=& \frac{1}{\sqrt{2}}[(\vert +y\rangle_A\vert -y\rangle_B + \vert
-y\rangle_A\vert +y\rangle_B)\vert +x\rangle_C \nonumber\\
&& + (\vert +y\rangle_A\vert +y\rangle_B + \vert -y\rangle_A\vert
-y\rangle_B)\vert -x\rangle_C,\label{GHZcorrelation}
\end{eqnarray}
where $\vert 0\rangle$ and $\vert 1\rangle$ are the two eigenvectors
of the operator $z$ (for example, the spin of a photon along the z
direction), and $\vert \pm x\rangle$ and $\vert \pm y\rangle$ are
those for the operators $x$ and $y$, i.e.,
\begin{eqnarray}
\vert +x\rangle &=& \frac{1}{\sqrt{2}}(\vert 0\rangle + \vert
1\rangle),\;\;\;\; \vert +y\rangle =\frac{1}{\sqrt{2}}(\vert
0\rangle + i\vert 1\rangle), \nonumber\\
\vert -x\rangle &=& \frac{1}{\sqrt{2}}(\vert 0\rangle - \vert
1\rangle),\;\;\;\; \vert -y\rangle =\frac{1}{\sqrt{2}}(\vert
0\rangle - i\vert 1\rangle).
\end{eqnarray}

The principle of the HBB99 QSS protocol \cite{HBB99} in the simple
case can be described as follows.

The boss Alice prepares a GHZ triplet in the state $\vert
\Psi\rangle_{ABC}=\frac{1}{\sqrt{2}}(\vert 000\rangle + \vert
111\rangle)_{ABC}$ and sends the photons $B$ and $C$ to her two
agents, Bob and Charlie, respectively, and keeps the photon $A$. All
the three parties choose randomly the measuring basis (MB) $x$ or
$y$ to measure their photons. They repeat their quantum
communication until they obtain an enough large set of outcomes for
generating the private key $K_A = K_B \oplus K_C$. Here $K_A$, $K_B$
and $K_C$ are the keys obtained by Alice, Bob and Charlie,
respectively. Then they publish the information about their MBs.
When they all choose the MB $x$ or two choose the MB $y$ and the
other one chooses the MB $x$, they retain the corresponding outcomes
for generating their private key, otherwise they discard them. From
the Eq.(\ref{GHZcorrelation}), one can see that their outcomes
retained are correlated. In other words, any two parties can deduce
the outcome of the other if they cooperate. For checking
eavesdropping, Alice chooses randomly a large subset of the outcomes
obtained by the three parties to analyze its error rate. That is,
Alice requires Bob and Charlie announce their outcomes on the
samples in public. If the error rate is lower than the threshold
value, they keep the outcomes for distilling the private key $K_A$.

As pointed out in Ref. \cite{KKI}, if QSS protocol is secure for the
dishonest agent, it is secure for any eavesdropper. In essence,
Alice requires the dishonest agent, say Bob publish his outcomes of
the samples in the process for checking eavesdropping in HBB99 QSS
protocol. Just his process makes Bob have the chance for hiding his
eavesdropping with cheating. Now let us introduce the
fake-signal-and-cheating attack on HBB99 protocol. It includes two
steps.

(1) Bob replaces the original photon $C$ with a fake photon $C'$.
That is, Bob intercepts the photon $C$ which is sent from Alice to
Charlie, and stores it. He sends a fake photon $C'$ in the Bell
state $\vert \phi^+\rangle_{B'C'}=\frac{1}{\sqrt{2}}(\vert 0\rangle
_{B'}\vert 0\rangle _{C'} + \vert 1\rangle _{B'}\vert 1\rangle
_{C'})$ to Charlie and keeps the photon $B'$. The process is
repeated until the parties stop the process for transmitting the
photons. Bob does not measure the photons $B$, $C$ and $B'$ until
Alice and Charlie publishes the MBs for their photons.

(2) Bob cheats the other parties when a GHZ triplet is chosen by
Alice as a sample for checking eavesdropping. In detail, Bob first
performs a Bell-state measurement on the photons $B'$ and $C$, and
then hides the difference between the state of the quantum system
composed of the photons $A$, $B$ and $C'$ and the original state
$\vert \Psi\rangle$ by publishing a fake information with a cheat.
\begin{eqnarray}
\vert \Psi\rangle\otimes \vert \phi^+\rangle &=&
\frac{1}{\sqrt{2}}(\vert 000\rangle + \vert
111\rangle)_{ABC}\nonumber\\
&& \otimes\frac{1}{\sqrt{2}}(\vert
00\rangle+\vert 11\rangle)_{B'C'}\nonumber\\
&=& \frac{1}{2\sqrt{2}}[(\vert 000\rangle + \vert
111\rangle)_{ABC'}\vert \phi^+\rangle_{CB'} \nonumber\\
&& +(\vert 000\rangle - \vert
111\rangle)_{ABC'}\vert \phi^-\rangle_{CB'} \nonumber\\
&&+ (\vert 001\rangle + \vert
110\rangle)_{ABC'}\vert \psi^+\rangle_{CB'} \nonumber\\
&& +(\vert 001\rangle - \vert 110\rangle)_{ABC'}\vert
\psi^-\rangle_{CB'}],\label{GHZswapping}
\end{eqnarray}
where
\begin{eqnarray}
\left\vert \psi ^{\pm}\right\rangle =\frac{1}{\sqrt{2}}(\left\vert
0\right\rangle\left\vert 1\right\rangle\pm\left\vert
1\right\rangle \left\vert 0\right\rangle ), \label{EPR12}\\
\left\vert \phi ^{\pm}\right\rangle =\frac{1}{\sqrt{2}}(\left\vert
0\right\rangle \left\vert 0\right\rangle \pm\left\vert
1\right\rangle \left\vert 1\right\rangle ). \label{EPR34}
\end{eqnarray}
That is, Bob can deduce the difference between the state of the
system $\vert \Psi\rangle_{ABC'}$ and the original state $\vert
\Psi\rangle_{ABC}$ according to his outcomes with the Bell-state
measurement on photons $B'$ and $C$. He can hide the difference by
publishing  a fake information with a cheat.
\begin{eqnarray}
\vert \Psi\rangle_1 &\equiv& \frac{1}{\sqrt{2}}(\vert 000\rangle -
\vert 111\rangle)_{ABC'}\nonumber\\
&=& \frac{1}{2}[(\vert +x\rangle\vert -x\rangle + \vert
-x\rangle\vert +x\rangle)_{AB}\vert +x\rangle_{C'} \nonumber\\
&& +(\vert +x\rangle\vert +x\rangle + \vert -x\rangle\vert
-x\rangle)_{AB}\vert -x\rangle)_{C'}] \nonumber\\
&=& \frac{1}{2}[(\vert +y\rangle\vert +y\rangle + \vert
-y\rangle\vert -y\rangle)_{AB}\vert +x\rangle_{C'} \nonumber\\
&& +(\vert +y\rangle\vert -y\rangle + \vert -y\rangle\vert
+y\rangle)_{AB}\vert -x\rangle_{C'}],\label{fakeGHZ1}\\
\vert \Psi\rangle_2 &\equiv& \frac{1}{\sqrt{2}}(\vert 001\rangle +
\vert 110\rangle)_{ABC'}\nonumber\\
&=& \frac{1}{2}[(\vert +x\rangle\vert +x\rangle + \vert
-x\rangle\vert -x\rangle)_{AB}\vert +x\rangle_{C'} \nonumber\\
&& -(\vert +x\rangle\vert -x\rangle + \vert -x\rangle\vert
+x\rangle)_{AB}\vert -x\rangle)_{C'}] \nonumber\\
&=& \frac{1}{2}[(\vert +y\rangle\vert -y\rangle + \vert
-y\rangle\vert +y\rangle)_{AB}\vert +x\rangle_{C'} \nonumber\\
&& -(\vert +y\rangle\vert +y\rangle + \vert -y\rangle\vert
-y\rangle)_{AB}\vert -x\rangle_{C'}],\label{fakeGHZ2}\\
\vert \Psi\rangle_3 &\equiv& \frac{1}{\sqrt{2}}(\vert 001\rangle -
\vert 110\rangle)_{ABC'}\nonumber\\
&=& \frac{1}{2}[(\vert +x\rangle\vert -x\rangle + \vert
-x\rangle\vert +x\rangle)_{AB}\vert +x\rangle_{C'} \nonumber\\
&& -(\vert +x\rangle\vert +x\rangle + \vert +x\rangle\vert
+x\rangle)_{AB}\vert -x\rangle)_{C'}] \nonumber\\
&=& \frac{1}{2}[(\vert +y\rangle\vert +y\rangle + \vert
-y\rangle\vert -y\rangle)_{AB}\vert +x\rangle_{C'} \nonumber\\
&& -(\vert +y\rangle\vert -y\rangle + \vert -y\rangle\vert
+y\rangle)_{AB}\vert -x\rangle_{C'}].\label{fakeGHZ3}
\end{eqnarray}

From Eqs.(\ref{GHZcorrelation}), (\ref{GHZswapping}) and
Eqs.(\ref{fakeGHZ1})-(\ref{fakeGHZ3}), one can see that Bob can hide
his eavesdropping by flipping his outcomes when he obtains the
outcomes of the Bell-state measurements $\vert \phi^-\rangle$ and
$\vert \psi^-\rangle$ whether he chooses the MB $x$ or $y$. As Alice
and Charlie cannot detect this eavesdropping, Bob can measure the
other photons transmitted from Alice to Charlie after all the
parties published their MBs for their photons. In this way, Bob can
obtain all the information about Charlie's outcomes with
single-photon measurements, i.e., $K_C$. With his key $K_B$, Bob can
read out the secret message sent by Alice without cooperating with
Charlie as he obtains the key $K_A$ fully and freely.

With this fake-signal-and-cheating attack strategy, the dishonest
agent can eavesdrop freely and fully  the information of the other
agent in KKI QSS protocol \cite{KKI}. The important difference
between HBB99 protocol \cite{HBB99} and KKI protocol \cite{KKI} is
that the quantum signal prepared by the boss Alice is a two-qubit
maximally entangled state, one of the four states $\{\vert
\psi^+\rangle_{BC}, \vert \phi^-\rangle_{BC},
\vert\Psi^+\rangle_{BC}, \vert \Phi^-\rangle_{BC}\}$. Here $\vert
\Psi^+\rangle=\frac{1}{\sqrt{2}}(\vert \phi^-\rangle +\vert
\psi^+\rangle)$ and $\vert \Phi^-\rangle=\frac{1}{\sqrt{2}}(\vert
\phi^-\rangle - \vert \psi^+\rangle)$. The photons $B$ and $C$ are
the particles sent from Alice to Bob and Charlie, respectively. In
the quantum communication, Bob and Charlie choose randomly the MBs
$x$ and $z$ to measure their photons received. After an enough set
of outcomes is produced, the three parties compare their MBs and
keeps those in which their MBs are correlated. They check the
eavesdropping by picking up some samples from their outcomes.

Same as the case that the dishonest agent eavesdrops the quantum
communication in HBB99 QSS protocol, Bob also prepares a fake
entangled state $\vert \phi^+\rangle_{B'C'}$ and replaces the
original photon $C$ sent from Alice to Bob with the photon $C'$. Bob
stores the photons $B$ and $C$. He measures them after the parties
publish their MBs for their photons. Suppose that the original state
of the sample chosen by the three parties is $\vert
\Psi'\rangle_{BC}=\alpha \vert 00\rangle_{BC} + \beta \vert
01\rangle_{BC} + \gamma\vert 10\rangle_{BC} + \delta\vert
11\rangle_{BC}$ ($\vert\alpha \vert^2 + \vert\beta \vert^2 +
\vert\gamma \vert^2 + \vert\delta \vert^2 =1 $). Bob can hide his
action with entanglement swapping and some a local unitary operation
on his photon $B$ as
\begin{widetext}
\begin{center}
\begin{eqnarray}
\vert \Psi'\rangle_{BC}\otimes \vert \phi^+\rangle_{B'C'}&=&(\alpha
\vert 00\rangle_{BC} + \beta \vert 01\rangle_{BC} + \gamma\vert
10\rangle_{BC} + \delta\vert 11\rangle_{BC})\otimes
\frac{1}{\sqrt{2}}(\vert
00\rangle + \vert 11\rangle)_{B'C'}\nonumber\\
&=& \frac{1}{2}[(\alpha \vert 00\rangle + \beta \vert 01\rangle +
\gamma\vert 10\rangle + \delta\vert 11\rangle_{BC'})\vert
\phi^+\rangle_{CB'} + (\alpha \vert 00\rangle - \beta \vert
01\rangle + \gamma\vert 10\rangle -
\delta\vert 11\rangle_{BC'} )\vert \phi^-\rangle_{CB'}\nonumber\\
&& + (\alpha \vert 01\rangle + \beta \vert 00\rangle + \gamma\vert
11\rangle + \delta\vert 10\rangle_{BC'})\psi^+\rangle_{CB'} +
(\alpha \vert 01\rangle - \beta \vert 00\rangle + \gamma\vert
11\rangle - \delta\vert 10\rangle_{BC'})\vert \psi^-\rangle_{CB'}].\nonumber 
\end{eqnarray}
\end{center}
\end{widetext}
That is, Bob can easily reconstruct the original entangled state
$\vert \Psi'\rangle_{BC}$ prepared by Alice when it is chosen for
checking eavesdropping. He need only perform the local unitary
operation $U_0$, $U_1$, $U_2$ or $U_3$ on his photon $B$ according
to the result of the Bell-state measurement on the photons $C$ and
$B'$ is $\vert \psi^-\rangle$, $\vert \psi^-\rangle$, $\vert
\phi^-\rangle$ or $\vert \phi^+\rangle$, respectively. Here
\begin{eqnarray}
U_{0}=\left\vert 0\right\rangle \left\langle 0\right\vert
+\left\vert 1\right\rangle \left\langle 1\right\vert, \,\,\,\,\,
U_{1}=\left\vert 0\right\rangle \left\langle 1\right\vert
-\left\vert 1\right\rangle \left\langle 0\right\vert, \nonumber\\
U_{2}=\left\vert 1\right\rangle \left\langle 0\right\vert
+\left\vert 0\right\rangle \left\langle 1\right\vert, \,\,\,\,\,\,
U_{3}=\left\vert 0\right\rangle \left\langle 0\right\vert
-\left\vert 1\right\rangle \left\langle 1\right\vert. \label{U}
\end{eqnarray}

In fact, almost all the QSS protocols
\cite{HBB99,KKI,Bandyopadhyay,Karimipour,Nascimento,Gottesman,cpyang,dengQSSentanglement,longqss,zhangPLA,zhanglm,yanpra},
in which the eavesdropping check is complete with the cooperation of
the dishonest one whether the protocol is based on one-way direction
quantum communication or two-way one, can be eavesdropped freely and
fully. The principle is same as those for eavesdropping the HBB99
\cite{HBB99} and the KKI \cite{KKI} QSS protocols with or without a
little modification.

For improving the security of this kind of QSS protocols
\cite{HBB99,KKI,Bandyopadhyay,Karimipour,Nascimento,Gottesman,cpyang,dengQSSentanglement,longqss,zhangPLA,zhanglm,yanpra},
Alice should provide a process for checking eavesdropping without
the cooperation of the dishonest agent. In a simple case, Alice can
insert some decoy photons in the quantum signal sent to each agent.
As the process for analyzing their error rate needs not the other
participants except for the one who received the decoy photons, the
action of the eavesdropping done by the dishonest one will be
detected. Let us use the HBB99 QSS protocol as an example for
describe the principle of the way for improving its security. In
detail, Alice prepares some decoy photons which are randomly in the
states $\{\vert +x\rangle, \vert -x\rangle, \vert +y\rangle, \vert
-y\rangle \}$ and then replaces the photons $B$ and $C$ in the GHZ
triplet $\vert \Psi\rangle_{ABC}$ with the decoy photons some a
probability $1-p$, similar to the QKD protocol in Ref. \cite{ABC}.
After the parties announce their MBs for their measurements on their
photons, Alice picks up the outcomes of the decoy photons as the
samples for checking eavesdropping. Even though half of the decoy
photons will be discarded as the MBs chosen by the agents and Alice
is not correlated, same as BB84 QKD protocol \cite{BB84}, the
process for checking eavesdropping  requires only one of the agents
cooperate with Alice to analyze the error rate of each subset of the
decoy photons, which does not give the chance to the dishonest agent
for eavesdropping the information of another agent freely.

The decoy photons can be prepared with another ideal single-photon
source. Also they can be prepared by measuring some of the photons
in the GHZ triplet $\vert \Psi\rangle$. That is, Alice can performs
the single-photon measurements on the photons $B$ and $C$ with the
MB $x$ or $y$, and the photon $A$ can be prepared to a state needed
by Alice. The advantage of the way for preparing the decoy photons
is that Alice need not resort to another ideal single-photon source.
The disadvantage is that Alice maybe waste some of the GHZ triplets.

In summary, we have present a fake-signal-and-cheating attack for
the dishonest agent to eavesdrop almost all the QSS protocols
\cite{HBB99,KKI,Bandyopadhyay,Karimipour,Nascimento,Gottesman,cpyang,dengQSSentanglement,longqss,zhangPLA,zhanglm,yanpra}
in which the process of checking eavesdropping has to resort to his
cooperation. With this attack, the dishonest one can obtain the
information of  another agent freely and fully. Moreover, we discuss
a possible way for improving the security of this kind of QSS
protocols.

This work is supported by the National Natural Science Foundation of
China under Grant Nos. 10447106, 10435020, 10254002 and A0325401,
and Beijing Education Committee under Grant No. XK100270454.

\end{document}